\documentclass[11pt,twoside]{article}
\usepackage{fancyhdr}
\usepackage[colorlinks,citecolor=blue,urlcolor=blue,linkcolor=blue,bookmarks=false]{hyperref}
\usepackage{amsfonts,epsfig,graphicx}
\usepackage{afterpage}
\usepackage{amsmath,amssymb,amsthm} 
\usepackage{fullpage}
\usepackage[T1]{fontenc} 
\usepackage{epsf} 
\usepackage{graphics} 
\usepackage{amsfonts,amsmath}
\usepackage[sort,numbers]{natbib} 
\usepackage{psfrag,xspace}
\usepackage{color,etoolbox}

\include{notations}

\usepackage{xspace,xcolor,enumerate}
\usepackage{amsmath,amsfonts,amssymb}
\usepackage{color,graphicx}
\usepackage{dsfont}
\usepackage{hyperref}
\usepackage{cancel}

\def\epsilon{\varepsilon}

\def\phi{\varphi}
\def\cal{\mathcal}

\renewcommand{\bar}{\overline}

\newcommand{\R}{{\mathbb R}}

\newcommand{\Vsymb}{\mathsf{Var}}
\newcommand{\Covsymb}{\mathsf{Cov}}

\DeclareMathOperator*{\VarOp}{\Vsymb}
\DeclareMathOperator*{\CovOp}{\Covsymb}

\newcommand{\var}[1]{\VarOp\left({#1}\right)}

\newcommand{\cov}[1]{\CovOp\left({#1}\right)}

\usepackage{nicefrac}

\newcommand{\norm}[1]{\ensuremath{\left\lVert #1 \right\rVert}}

\newcommand{\ip}[1]{\left\langle #1 \right\rangle}

\def\bfx {{\bf x}}
\def\bfy {{\bf y}}

\def\bfX {{\bf X}}
\def\bfY {{\bf Y}}

\def\bfu{{\bf u}}
\def\bfv{{\bf v}}

\newfont{\inhead}{eufm10 scaled\magstep1}

\newcommand{\calR}{{\cal R}}
\newcommand{\calX}{{\cal X}}
\newcommand{\calY}{{\cal Y}}

\newcommand{\maximize}{\operatornamewithlimits{\text{maximize}}}

\newcommand{\inparen}[1]{\left(#1\right)}             

\newlength{\tpush}
\newcommand{\titlebox}[5]
{
   \noindent
   \begin{center}
   \framebox{
      \vbox{\vspace{2mm}
    \hbox to 6.28in { { #1 \hfill #2} }
       \vspace{4.2mm}
       \hbox to 6.28in { {\sc \large \hfill #3 \hfill} }
       \vspace{2.2mm}
       \hbox to 6.28in { {#4 \hfill #5} }
      \vspace{2mm}}
   }
   \end{center}
   \vspace*{4mm}
}

\usepackage{subcaption}
\newcommand{\dx}{d_\calX}
\newcommand{\dy}{d_\calY}
\newcommand{\Dx}{{\bf D}^\calX}
\newcommand{\Dy}{{\bf D}^\calY}

\begin{document}



\begin{center} {\Large{\bf{Local White Matter Architecture Defines \\
\vspace{0.2cm}
Functional Brain Dynamics}}}
\\

\vspace*{.3in}

{\large{
\begin{center}
Yo Joong Choe{\let\thefootnote\relax\footnote{{$^\dagger$Y. C. is currently on leave from Carnegie Mellon University, where most of this work was done. \indent\indent\indent\indent\;\; 
\textcircled{c} 2018 IEEE. Personal use of this material is permitted. Permission from IEEE must be obtained for all other uses, in any current or future media, including reprinting/republishing this material for advertising or promotional purposes, creating new collective works, for resale or redistribution to servers or lists, or reuse of any copyrighted component of this work in other works. 
}}}$^\dagger$ ~~~~~ Sivaraman Balakrishnan$^\ddagger$ ~~~~~ Aarti Singh$^{\ddagger}$ \\
\vspace{.2cm}
Jean M. Vettel$^{\ast}$ ~~~~~ Timothy Verstynen$^{\ddagger}$ \\
\end{center}

\addtocounter{footnote}{-1}

\vspace*{.1in}

\begin{tabular}{c}
$^\dagger$Kakao\\
$^\ddagger$Carnegie Mellon University \\
$^\ast$U.S. Army Research Laboratory\\
\end{tabular}

\vspace*{.2in}

\begin{tabular}{c}
{\texttt{yj.c@kakaocorp.com, siva@stat.cmu.edu, aarti@cs.cmu.edu }} \\
 {\texttt{jean.m.vettel.civ@mail.mil, timothyv@andrew.cmu.edu}}
\end{tabular}
}}

\vspace*{.2in}

\today
\vspace*{.2in}





\begin{abstract}

Large bundles of myelinated axons, called white matter, anatomically connect disparate brain regions together and compose the structural core of the human connectome. We recently proposed a method of measuring the local integrity along the length of each white matter fascicle, termed the local connectome \cite{yeh2016connectometry}. If communication efficiency is fundamentally constrained by the integrity along the entire length of a white matter bundle \cite{pajevic2014role}, then variability in the functional dynamics of brain networks should be associated with variability in the local connectome. 
We test this prediction using two statistical approaches that are capable of handling the high dimensionality of data.
First, by performing statistical inference on distance-based correlations, we show that similarity in the local connectome between individuals is significantly correlated with similarity in their patterns of functional connectivity.
Second, by employing variable selection using sparse canonical correlation analysis and cross-validation, we show that segments of the local connectome are predictive of certain patterns of functional brain dynamics.
These results are consistent with the hypothesis that structural variability along axon bundles constrains communication between disparate brain regions.

\end{abstract}

\end{center}

\section{Introduction}
\label{intro}

The function of macroscopic neural networks is constrained by the integrity of structural connections between disparate regions. This form of long-distance (i.e., centimeters) communication relies on dense bundles of axons that are known as white matter \cite{passingham2002anatomical}. To prevent degradation of action potentials across long distances, these fiber bundles are supported by the myelin sheath, non-neuronal glial cells that insulate axons and facilitate communication along the fascicle. As a result, the integrity of the myelin sheath is critical for synchronizing information transmission between distal brain areas \cite{pajevic2014role}, fostering the ability of these networks to adapt over time \cite{fields2015new}. Thus, variability in the myelin sheath, as well as other cellular support mechanisms, would contribute to variability in functional coherence across the circuit.

To study the integrity of structural connectivity, we recently introduced the concept of the local connectome. This is defined as the pattern of fiber systems (i.e., number of fibers, orientation, and size) within a voxel, as well as immediate connectivity between adjacent voxels, that can be quantified using diffusion MRI (dMRI) by measuring the fiber-wise density of microscopic water diffusion within a voxel \cite{yeh2016connectometry}. The collection of these multi-fiber diffusion density measurements within all white matter voxels is termed the local connectome fingerprint (LCF). The LCF is a high-dimensional feature vector that describes the unique configuration of the structural connectome along the segments of white matter pathways \cite{yeh2016quantifying}. Thus, the LCF provides a diffusion-informed measure along the fascicles that supports inter-regional communication, rather than determining the start and end positions of a particular fiber bundle.

Since the LCF measures the local integrity along white matter bundles that connect regions across the entire brain, it reflects the overall communication capacity of the brain \cite{pajevic2014role}. Hence, we expect to see that variations in the LCF should also correlate with those in the dynamics of brain networks, measured by connectivity patterns in the resting-state functional MRI (fMRI). 
To formally validate this intuition, we employ statistical approaches to examine the following hypotheses: \\
\vspace{0.1cm}

\textbf{Hypothesis 1}\quad Similarity in the LCF, between individuals, is associated with similarity in their functional connectivity patterns measured with resting-state fMRI. \\
\vspace{0.1cm}

\textbf{Hypothesis 2}\quad Variability in specific segments of the LCF is associated with patterns of functional connectivity in specific circuits. 

\section{Materials and Methods}
\label{sec:materials}
We summarize our abbreviations and notation in Table \ref{tab:notation}.

\begin{table}[t]
\centering
\caption{Summary of Abbreviations and Notation}
\label{tab:notation}       
\begin{tabular}{cc}
\hline\noalign{\smallskip}
Notation & Definition  \\
\noalign{\smallskip}\hline\noalign{\smallskip}
LCF & Local connectome fingerprint \\
FCG & Functional correlation graph \\
HCP & The Human Connectome Project (dataset) \\
$n$ & Number of subjects (793) \\
$p$ & Dimension of LCF vectors (433,386) \\
$q$ & Dimension of FCG vectors (195,625) \\
$\bfx_i$ & $p$-dimensional LCF vector of subject $i$ \\
$\bfy_i$ & $q$-dimensional FCG vector of subject $i$ \\
$\dx$ & The scaled Euclidean distance between LCFs \eqref{eqn:dx} \\
$\dy$ & The Pearson correlation distance between FCGs \eqref{eqn:dy} \\
$\Dx$ & Scaled Euclidean distance matrix between $n$ LCFs \\
$\Dy$ & Pearson correlation distance matrix between $n$ FCGs \\
$\bfX$ & $n \times p$ matrix containing the $n$ LCFs as rows \\
$\bfY$ & $n \times q$ matrix containing the $n$ FCGs as rows \\
$\norm{\cdot}_1$, $\norm{\cdot}_2$ & The $\ell_1$- and $\ell_2$-norm of a real-valued vector \\
$\mathbb{S}_n$ & The permutation group on $\{1, \dotsc, n\}$ \\
\noalign{\smallskip}\hline
\end{tabular}
\end{table}

\subsection{Data Acquisition}
\label{sec:data}

\subsubsection{Participants}
We used publicly available dMRI and fMRI data from the S900 (2015) release of the Human Connectome Project (HCP) \cite{van2013wu}, acquired by Washington University in St. Louis and the University of Minnesota. Out of the 900 participants released, 841 genetically unrelated participants (370 male, ages 22-37, mean age 28.76) had viable dMRI datasets. Among them, $n=793$ participants had at least one viable resting-state fMRI measurement. Our analysis was restricted to this subsample. All data collection procedures were approved by the institutional review boards at Washington University in St. Louis and the University of Minnesota. The post hoc data analysis was approved as exempt by the institutional review board at Carnegie Mellon University, in accordance with 45 CFR 46.101(b)(4) (IRB Protocol Number: HS14-139). 

\subsubsection{Diffusion MRI Acquisition} 
The dMRI data were acquired on a Siemens 3T Skyra scanner using a 2D spin-echo single-shot multiband EPI sequence with a multi-band factor of 3 and monopolar gradient pulse. The spatial resolution was 1.25 mm isotropic (TR = 5500 ms, TE = 89.50 ms). The b-values were 1000, 2000, and 3000 s/mm2 . The total number of diffusion sampling directions was 90 for each of the three shells in addition to 6 b0 images. The total scanning time was approximately 55 minutes.

\subsubsection{LCF Reconstruction} 
\begin{figure}[t]
\centerline{\includegraphics[width=15cm]{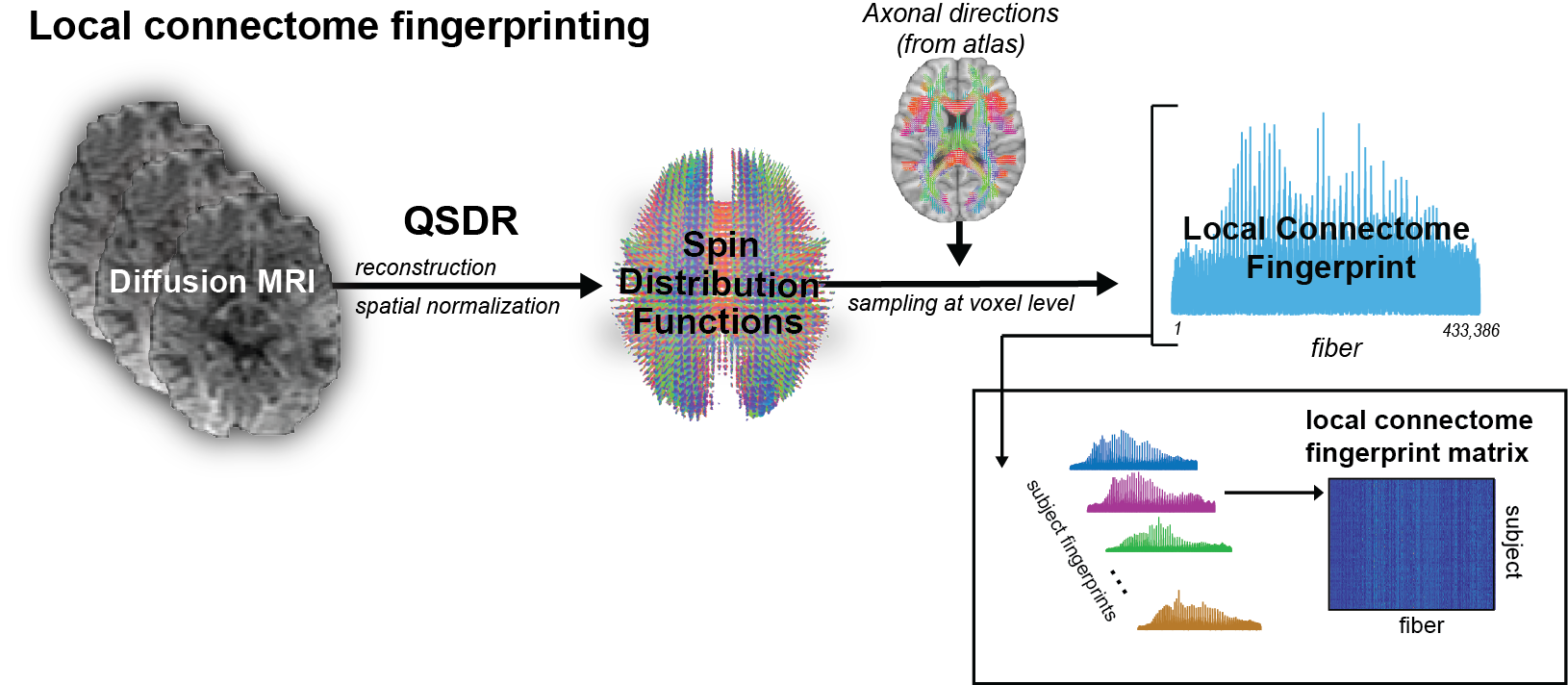}}
\caption{Pipeline for generating LCFs. See \cite{yeh2016quantifying} for details.}
\label{fig:lcf}
\end{figure}

An outline of the pipeline for generating LCFs is shown in Fig. \ref{fig:lcf}. The dMRI data for each subject was reconstructed in a common stereotaxic space using q-space diffeomorphic reconstruction (QSDR) \cite{yeh2011ntu}, a nonlinear registration approach that directly reconstructs water diffusion density patterns into a common stereotaxic space at 1 mm resolution. The LCF reconstruction was conducted using DSI Studio (http://dsi-studio.labsolver.org), an open-source diffusion MRI analysis tool for connectome analysis. To compute the LCF, the axonal direction in each voxel was derived from the HCP dataset, and all of the data and source code for this analysis are publicly available on the same website.

A spin distribution function (SDF) sampling framework was used to provide a consistent set of directions to sample the magnitude of SDFs along axonal directions in the cerebral white matter. Since each voxel may have more than one axonal direction, multiple measurements were extracted from the SDF for voxels that contained crossing fibers, while a single measurement was extracted for voxels with fibers in a single direction. The appropriate number of density measurements from each voxel was sampled by the left-posterior-superior voxel order and compiled into a sequence of scalar values. Gray matter was excluded using the ICBM-152 white matter mask (MacConnell Brain Imaging Centre, McGill University, Canada). The cerebellum was also excluded due to different slice coverage in cerebellum across participants. Since the density measurement has arbitrary units, the LCF was scaled to make the variance equal to 1 \cite{yeh2016quantifying}. For each subject $i = 1, \dotsc, n$, we denote this high-dimensional LCF of the $i$th subject, across $p = 433,386$ sampled directions, as $\bfx_i \in \R^p$. The collection of all $n$ LCFs are compactly represented as a data matrix $\bfX = [\bfx_1, \dotsc, \bfx_n]^T \in \R^{n \times p}$ with each LCF as a row vector. \\

\subsubsection{Functional MRI Acquisition \& Processing}
We analyzed the minimally processed resting-state fMRI data acquired as part of the Human Connectome Project (HCP) \cite{glasser2013minimal,van2013wu} which used a multi-band gradient echo-planar imaging protocol (see \cite{moeller2010multiband} for details on aquisition parameters). The dataset contains volumetric NIFTI data for resting-state fMRI scans (14 minutes each), motion parameters, and physiological data. Only data for the first resting-state scan collected at the A-P phase encoding direction were used for analyses. Using these measurements, we computed the average BOLD (blood-oxygen-level dependent) signals at each of the 626 regions of interest (ROIs) \cite{hermundstad2013structural} and regressed out the linear effects of the noise terms via ordinary least-squares (OLS). The 16 noise terms include the global signal, 12 motion parameters (6 estimates from a rigid-body transformation to the SBRef image acquired at the start of each scan; 6 temporal derivatives of these estimates), and the top-3 principal component projections of the voxel-level white matter signals (measured at each of 2,258 voxels and 840 seconds). The resulting residual terms were then filtered by a first-order Butterworth bandpass filter \cite{butterworth1930theory} between frequencies 0.08Hz and 0.15Hz.

\subsubsection{Functional Correlation Graph Construction}
For each subject, given a pre-processed time series (840s; 1Hz) at each ROI, we computed the functional correlation graph (FCG), alternatively called the functional connectome fingerprint in \cite{finn2015functional}, by computing the Pearson correlation between time series at every pair of ROIs. 
For each subject $i = 1, \dotsc, n$, we use $\bfy_i \in \R^q$ to denote the vector of these $q = {626 \choose 2} = 195,625$ Pearson correlations, which we collectively refer to as the $i$th FCG. The collection of all $n$ FCGs are compactly represented as a data matrix $\bfY = [\bfy_1, \dotsc, \bfy_n]^T \in \R^{n \times q}$ with each FCG as a row vector.

\subsection{Statistical Inference of Distance-based Correlations}
\label{sec:dcorr}

Our first goal is to test whether there is a statistically significant relationship between LCFs and FCGs. However, because both the structural and functional feature vectors are high-dimensional, fully multivariate statistical tests of dependence are intractable and uninterpretable. This means that we need to find a way to effectively reduce the dimensionality of each feature vector.  

For each pair of subjects, we first compute the pairwise distance between their feature vectors. This gives us one distance matrix between their structural features (LCFs) and another between their functional features (FCGs). Then, we measure the correlation between the resulting pair of structural and functional distance matrices. 

Our hypothesis states that if two subjects have similar LCFs, then they are more likely to also have similar FCGs. This hypothesis derives from previous research that found (a) similar LCFs imply genetic similarity \cite{yeh2016quantifying} and (b) identical FCGs imply that the two graphs most likely come from the same individual \cite{finn2015functional}. By formally defining a notion of similarity, it is possible to derive distribution-free statistical inference methods that can test whether the two high-dimensional feature vectors are correlated or not. This approach overcomes the high dimensionality while being statistically and theoretically rigorous.

\subsubsection{Choice of Distance Metrics}
In \cite{yeh2016quantifying}, Yeh et al. establish that LCFs are highly specific to each individual. More precisely, they show that the Euclidean distance between any pair of LCFs effectively captures the genetic (and temporal) difference between the two measurements, achieving 100\% accuracy across 17,398 leave-one-out identification tasks. Therefore, to quantify individual variability in structural features, we use the Euclidean distance, scaled by the number of features as in \cite{yeh2016quantifying}:
\begin{equation}\label{eqn:dx}
\dx(\bfx, \bfx') = \frac{1}{p} \norm{\bfx - \bfx'}_2 = \frac{1}{p}\sqrt{\sum_{k=1}^p (x_k - x_k')^2}
\end{equation}
To estimate distance between functional features, we follow the approach that Finn et al. \cite{finn2015functional} used on FCGs of the Q2-released version of the HCP dataset. They successfully predicted identity with 92.9-94.4\% test set accuracy using the Pearson correlation, and the accuracy increased to 98-99\% when comparing specific sub-networks (the medial frontal network and the frontoparietal network). Since our goal is to capture individual variability, not maximize prediction accuracy, we use the Pearson correlation distance on the entire FCG:
\begin{align}\label{eqn:dy}
\dy(\bfy, \bfy') &= 1 - \rho(\bfy, \bfy') \nonumber \\
&= 1 - \frac{\sum_{l=1}^q(y_l - \bar{\bfy})(y_l' - \bar{\bfy'})}{\sqrt{\sum_{l=1}^q(y_l-\bar{\bfy})^2}\sqrt{\sum_{l=1}^q(y_l'-\bar{\bfy'})^2}}
\end{align}
where $\rho$ denotes the Pearson correlation and $\bar{\bfy} = \frac{1}{q}\sum_{l=1}^l y_l$ denotes the mean of all entries in the vector $\bfy$. 
We note that $\dy$ is not a proper distance metric in the mathematical sense, because it does not satisfy positive definiteness or triangle inequality. It is nevertheless nonnegative, symmetric, and is exactly zero when the two inputs are identical (it is also zero when two inputs are scalar multiples of each other). 

Given these choices of metrics, we can represent all such distances on our data compactly in two $n \times n$ distance matrices, $\Dx \in \R^{n\times n}$ and $\Dy \in \R^{n\times n}$, such that $\Dx_{ij} = \dx(\bfx_i, \bfx_j)$ and $\Dy_{ij} = \dy(\bfy_i, \bfy_j)$.

\subsubsection{Setting Up a Valid Hypothesis Test}
In general, it is highly nontrivial to set up a proper statistical test comparing distance matrices, because the entries of each distance matrix are not independent from each other. Intuitively, for any pair of subjects $i$ and $j$, the distance between the $i$th and $j$th feature vectors is correlated with the distance between the $i$th feature vector and any other feature vector. 
Thus, standard statistical approaches that rely on the i.i.d. assumption will \emph{not} give valid results if they are na\"ively applied to distance matrices. 

We instead use the distance matrices to construct null and alternative hypotheses and derive proper statistical inference strategies. Given independent copies of random vectors $(\bfx, \bfy) \sim P_{\calX\calY}$, where $P_{\calX\calY}$ is the joint distribution of $\bfx$ and $\bfy$, we test
\begin{equation}\label{eqn:test}
H_0: \calR(\bfx, \bfy) = 0 \quad\text{and}\quad  H_1: \calR(\bfx, \bfy) > 0
\end{equation}
where 
\begin{align}\label{eqn:dcorr}
\calR(\bfx, \bfy) &= \rho(\dx(\bfx, \bfx'), \dy(\bfy, \bfy')) \nonumber \\
&= \frac{\cov{\dx(\bfx, \bfx'), \dy(\bfy, \bfy')}}{\sqrt{\var{\dx(\bfx, \bfx')}}\sqrt{\var{\dy(\bfy, \bfy')}}}
\end{align}
In short, $\calR$ is the Pearson correlation between the two random distances, each of which is a function of two independent and identically distributed random variables. In our approach, the null hypothesis states that the Euclidean distance between the LCFs of two subjects is uncorrelated with the correlation distance between their FCGs. The alternative hypothesis states that the two distances are in fact positively correlated. 
Note that it is natural to consider a one-sided hypothesis here because we know that both distances are likely to increase as two subjects become more genetically distant \cite{yeh2016quantifying, finn2015functional}.

While there are no known parametric statistical tests corresponding to \eqref{eqn:test}, we can extend the permutation test of Pearson correlation in standard linear regression to our case. 
Given the structural and functional distance matrices $\Dx \in \R^{n\times n}$ and $\Dy \in \R^{n\times n}$, let $\bar\Dx = \frac{1}{{n \choose 2}} \sum_{i<j} \Dx_{ij}$ and $\bar\Dy = \frac{1}{{n \choose 2}} \sum_{i<j} \Dy_{ij}$, where $\Sigma_{i<j}$ denotes the double sum $\sum_{i=1}^n\sum_{j=i+1}^n$.
Then, the sample test statistic for \eqref{eqn:test} is given by
\begin{align}\label{eqn:sample_dcorr}
\widehat\calR_n(\bfX, \bfY) = 
\frac{ \sum_{i<j}\inparen{\Dx_{ij} - \bar\Dx}\inparen{\Dy_{ij} - \bar\Dy} }
{ \sqrt{\sum_{i<j}\inparen{\Dx_{ij} -
\bar\Dx}^2}\sqrt{\sum_{i<j}\inparen{\Dy_{ij} - \bar\Dy}^2} }
\end{align}

Given \eqref{eqn:sample_dcorr}, a permutation test can be devised by randomly shuffling one of the feature vectors (say $\bfx$, without loss of generality) among the $n$ subjects. This corresponds to permuting the rows of the data matrix $\bfX \in \R^{n \times p}$.
Mathematically, for a random permutation $\sigma \in \mathbb{S}_n$ of $n$ elements, the empirical distribution of \emph{permuted} correlations
\begin{align}\label{eqn:sample_dcorr_perm}
\widehat\calR_{n, \sigma}(\bfX, \bfY) = 
\frac{ \sum_{i<j}\inparen{\Dx_{\sigma(i),\sigma(j)} - \bar\Dx}\inparen{\Dy_{ij} - \bar\Dy} }
{ \sqrt{\sum_{i<j}\inparen{\Dx_{ij} -
\bar\Dx}^2}\sqrt{\sum_{i<j}\inparen{\Dy_{ij} - \bar\Dy}^2} }
\end{align}
estimates the null distribution of $\calR(\bfx, \bfy)$. 
If the sample correlation \eqref{eqn:sample_dcorr} deviates from this null distribution significantly, then we can reject the null hypothesis of the test in \eqref{eqn:test}. 

This test can be viewed as a variant of the Mantel test \cite{mantel1967detection}, which jointly permutes both features among the $n$ subjects to test the same statistic. Yet, because our version does not permute the feature dimension, it does not introduce unintended bias coming from spatial correlations \cite{guillot2013dismantling}.

Note that a nonzero correlation will imply statistical dependence, but not the other way around. When we take $\dy$ to be the Euclidean distance instead of the correlation distance, however, we obtain distance correlation (dCor) \cite{szekely2007measuring}, where a zero value implies statistical independence. We will consider the statistical test \eqref{eqn:test} both when $\dy$ is the correlation distance and when $\dy$ is the Euclidean distance. In the latter case, we use the unbiased version of the statistic that leads to a $t$-test \cite{szekely2013distance}.\footnote{We implement the unbiased dCor $t$-test \cite{szekely2013distance} by (substantially) modifying the MATLAB implementation found in \url{http://mathworks.com/matlabcentral/fileexchange/39905-distance-correlation}.}

\subsubsection{Constructing a Valid Confidence Interval with Subsampling}
The permutation test is nonparametric, but it does not readily yield confidence intervals unless a stronger assumption (and tedious computation) is made \cite{john1983significance,garthwaite1996confidence}.
Subsampling \cite{politis1999subsampling} is an alternative approach to statistical inference that makes less assumptions and gives a confidence interval as its outcome. 
It estimates the true distribution of $\calR(\bfx, \bfy)$ by computing the empirical version of the statistic many times on different random subsamples of the full data. 

Subsampling notably differs from the more standard bootstrapping because it samples \emph{without} replacement and only samples a fraction of the $n$ data points. 
The first difference is crucial in our scenario, because any duplicate sample from bootstrapping will zero out entries of $\Dx$ and $\Dy$ and thus lead to a biased (higher) estimate of $\calR(\bfx, \bfy)$. 

\subsection{High-Dimensional Canonical Correlation Analysis with Cross-Validation}
\label{sec:cca}

While statistical inference of the distance-based correlation will provide some insights to the structure-function relationship, this measure of correlation aggregated over so many features may not be as intuitive or informative. 
In search of more detailed and interpretable relationships between the two sets of features, we attempt to find small subsets of the LCF that are predictive of small subsets of the FCG \emph{on a held-out set}. 

\subsubsection{Canonical Correlation Analysis}
For a pair of random vectors, canonical correlation analysis (CCA) \cite{hotelling1936relations} finds a pair of linear transformations (``alignments'') onto the same Euclidean space such that the projections are the most correlated. 
Assuming \emph{centered} data $\bfX = [\bfx_1, \dotsc, \bfx_n]^T \in \R^{n \times p}$ and $\bfY = [\bfy_1, \dotsc, \bfy_n]^T \in \R^{n \times q}$, CCA solves the following biconvex constrained optimization problem:
\begin{align}\label{eqn:cca}
\maximize_{\bfu \in \R^p, \bfv \in \R^q} &\quad \ip{\bfX\bfu, \bfY\bfv} \\
\text{subject to} &\quad \norm{\bfX\bfu}_2^2 \leq 1 \nonumber \\ 
				  &\quad \norm{\bfY\bfv}_2^2 \leq 1 \nonumber
\end{align}
The objective is often written alternatively as $\bfu^T \widehat\Sigma_{\bfx\bfy} \bfv$, up to a $\frac{1}{n}$ constant, where $\widehat\Sigma_{\bfx\bfy} = \frac{1}{n} \bfX^T\bfY$ is the empirical cross-covariance matrix. 
When the columns of $\bfX$ and $\bfY$ are further standardized, the solution to this biconvex problem is given by the left and right singular vectors of the empirical cross-covariance matrix $\widehat\Sigma_{\bfx\bfy}$ that correspond to its largest singular value. 

Intuitively, CCA captures the directions in $\calX$ and $\calY$ that explain the largest cross-correlation. If we assume that $\bfX$ and $\bfY$ indeed have some correlation structures, then CCA will find the linear transformations that recover such structures.

\subsubsection{Sparse CCA}

In high dimensions, i.e. when the data dimensions $p$ and $q$ are large compared to the sample size $n$, the estimate $\widehat\Sigma_{\bfx\bfy}$ of the true cross-covariance is no longer consistent unless more structural assumptions are made \cite{bao2014canonical,johnstone2008multivariate}. 
It is also considered a more difficult problem than sparse PCA \cite{chen2013sparse,gao2015minimax}, which itself is considered challenging due to the poor behavior of the sample covariance matrix as an estimate \cite{johnstone2001distribution}.
To obtain a reliable estimate of the high-dimensional cross-correlation structure, we assume that there are interesting low-dimensional correlation structures between subsets of the structural and functional features. This allows us to focus on a sparse subset of each set of features that are the most correlated to one another.

A popular approach to finding sparse subsets of features is to use $\ell^1$-regularization. In our case, we add an $\ell_1$-penalty to the alignment vectors in \eqref{eqn:cca}:
\begin{align}\label{eqn:l1}
\norm{\bfu}_1 \leq c_1 \quad\text{and}\quad \norm{\bfv}_1 \leq c_2
\end{align}
where $c_1, c_2 > 0$ are sparsity parameters.
The $\ell_1$-penalty, most commonly used in the Lasso \cite{tibshirani1996regression}, performs variable selection by forcing some of the entries to be precisely zero when the sparsity parameters are sufficiently small. 
A penalized version of CCA that combines \eqref{eqn:cca} and \eqref{eqn:l1} has been called sparse CCA in the literature, and an alternating convex optimization algorithm can be used to find a sparse solution \cite{witten2009penalized,parkhomenko2007genome}. 


Yet, the $\ell_1$-penalty alone is not sufficient for effective variable selection in our setting. One reason is that both the LCF and the FCG naturally contain interesting correlation structures within their entries, while $\ell_1$-regularization tends to select only one entry from a correlated group \cite{zou2005regularization}.
Another reason is that $\ell_1$-penalized CCA from \eqref{eqn:cca} and \eqref{eqn:l1} is not strictly biconvex in high dimensions, so that the optimization problem can be unstable.
Both of these issues can be alleviated by further including an $\ell_2$-penalty:
\begin{align}
\norm{\bfu}_2 \leq d_1 \quad\text{and}\quad \norm{\bfv}_2 \leq d_2
\end{align}
with constants $d_1, d_2 > 0$.\footnote{For simplicity, we fix these constants to be $1$ in our analysis.}
The resulting optimization problem can be viewed as the elastic net \cite{zou2005regularization} applied to CCA. 
It is now a strictly biconvex problem, and we can find a feasible solution efficiently by alternately applying existing convex optimization solvers. 
We note that, in general, there is no known algorithm for this biconvex problem that guarantees a globally optimal solution \cite{witten2009penalized}.
For our analysis, we use the MATLAB implementation from \cite{chen2012structured} that is freely available online.\footnote{\url{http://people.stern.nyu.edu/xchen3/Code/groupCCA.zip}}

\subsubsection{$k$-Fold Cross-Validation}\label{sec:cv}
For sparse CCA, we use $k$-fold cross-validation to find the set of sparsity parameters that give the highest canonical correlations between subsets of the LCFs and the FCGs.

Specifically, using $k=5$, we first split the $n$ subjects into training and test sets with the ratio of 5 to 1. Then, we randomly partition the training set (size $\lceil{5/6}\rceil n$) into 5 equally sized subsamples, fit sparse CCA with each candidate set of sparsity parameters to 4 of the subsamples, and use the fitted alignment vectors $\bfu$ and $\bfv$ to align the feature vectors from the unused subsample (i.e. the validation set). 
The resulting canonical correlation on the validation set can be viewed as an estimate of canonical correlation on unseen data. 
By leaving out each of the 5 subsamples in the previous step, we obtain 5 such estimates of the canonical correlation, and the average of these 5 estimates can be used to validate the performance of the candidate set of sparsity parameters. 
After these steps, we choose the set of sparsity parameters that give the largest average canonical correlation on the validation set.

The resulting alignment vectors can transform unseen feature vectors coming from the same distribution as our dataset, so that the LCFs are the most correlated to the FCGs in the transformed space. 
The final performance of these alignment vectors is measured by the correlation between the alignments of the test set, which was unused throughout the cross-validation steps.

\section{Results}

\subsection{Exploratory Analysis}

\begin{figure}[t]
	\centering
    \includegraphics[width=0.78\textwidth]{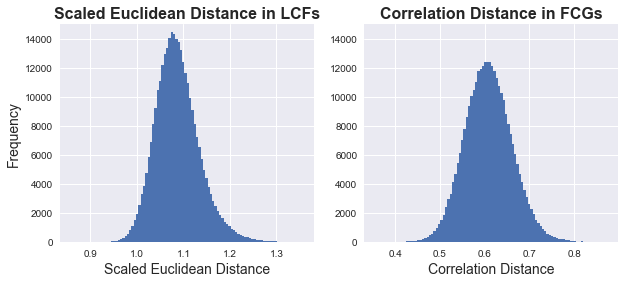}
	\caption{Pairwise distances between pairs of subjects' LCFs (left) and FCGs (right).}
	\label{fig:pdist}
\end{figure}

We first present exploratory analysis results for the inter-subject distances in LCFs and FCGs. 
Fig. \ref{fig:pdist} shows that the feature distances between different subjects appear substantially distant from zero. This in part reproduces the results from \cite{yeh2016quantifying} and \cite{finn2015functional}, in which it is shown that the distances between different individuals are significantly greater than those between the same subjects. 
This justifies our choice of distances \eqref{eqn:dx} and \eqref{eqn:dy} for the permutation test as well as the subsampling-based confidence interval.

\subsection{Statistical Inference}

In Table \ref{tbl:stat}, we summarize our results from the permutation test, the dCor $t$-test, and the subsampling-based confidence interval. Significance levels are marked with \textsuperscript{*} ($p < .05$), \textsuperscript{**} ($p < .01$), and \textsuperscript{***} ($p < .001$). Significant confidence intervals are marked with \textsuperscript{+}. 
We used $100,000$ random permutations for the hypothesis tests as well as $100,000$ subsamples for the confidence interval construction. Subsampling ratio was chosen as 0.135, following the procedure in \cite{bickel2008choice}.

\begin{table}[t]
  \linespread{1.5}
  \caption{Summary of Statistical Inference Results ($n=793$)}\label{tbl:stat}
  \centering
  \scalebox{1.0}{
  \begin{tabular}{ c | c | c | c }
    \hline
    {\bf Method} & {\bf Correlation} & {\bf Result Type} & {\bf Result} \\
    \hline
    {\bf Permutation} \eqref{eqn:sample_dcorr_perm} & {0.120} & {$p$-value} & {\bf $<$ 0.001\textsuperscript{***}}  \\
    {\bf dCor $t$-test} \cite{szekely2013distance} & 0.252 & $p$-value & {\bf $<$ 0.001\textsuperscript{***}} \\ \hline
    {\bf Subsampling} & {0.120} & {95\% conf. int.} & {\bf (0.098, 0.141)\textsuperscript{+}} \\
    \hline
  \end{tabular}
  }
\end{table}

Using a significance level of $\alpha = 0.05$, we find from the permutation test that there is indeed a statistically significant correlation between the Euclidean distances in LCFs and the correlation distances in FCGs. The dCor $t$-test of independence confirms that the two sets of features are statistically dependent, despite the fact that the test makes strong assumptions. 
Further, because the 95\% confidence interval does not include zero, we conclude that the correlation between LCF distances and FCG distances is statistically significant. 

Each of these results indicate that the similarity in the local connectome between individuals is significantly correlated to the similarity in their functional connectivity patterns. Specifically, our results show that if two individuals have similar local white matter architectures, they are also more likely to have similar functional brain dynamics.

Note that the correlation value for permutation test and subsampling are indeed identical, because they both compute exactly \eqref{eqn:sample_dcorr}. The value in dCor $t$-test \cite{szekely2013distance} differs, however, not only because the distance metric is changed to the Euclidean distances but because the test uses an unbiased estimate of the (Euclidean distance-based) statistic. While conceptually similar, the two computed values are estimates of different statistics and thus cannot be compared directly.

\begin{figure}[t]
	\centering
    \includegraphics[width=0.78\textwidth]{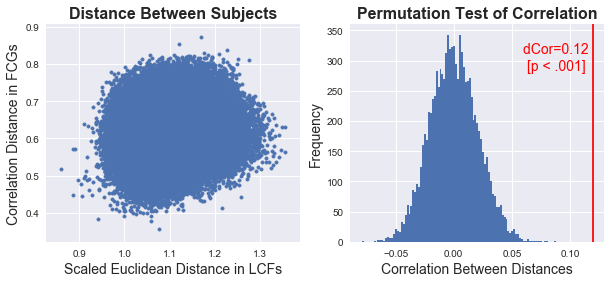}
	\caption{(Left) Scatterplot of all pairwise distances between the LCFs ($x$-axis) and between the FCGs ($y$-axis). (Right) Simulated null distribution using 10,000 random permutations of subjects \eqref{eqn:sample_dcorr_perm}. Red vertical line indicates the correlation on the actual dataset \eqref{eqn:sample_dcorr}. The $p$-value is the proportion of random correlations that fall on the right side of the red vertical line.}
	\label{fig:permtest}
\end{figure}
\begin{figure}[t]
	\centering
	\includegraphics[width=0.78\textwidth]{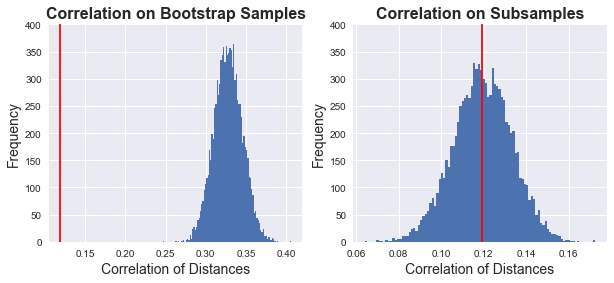}
	\caption{Histogram of linear correlations for 10,000 bootstrap samples (left) and 10,000 subsamples (right) of the HCP ($n=793$) dataset. Red vertical line indicates the correlation on the actual dataset (0.120).}
	\label{fig:why_subsampling}
\end{figure}

Fig. \ref{fig:permtest} visualizes the result from our permutation tests. 
On the left, we plot the structural and functional pairwise distances in a scatterplot to explore the overall trend. The scatterplot suggests that there is a positive trend between the pairwise distances in the structural and functional features. 
On the right, the permutation test shows that the correlation on real data is on the far-right tail of the correlation on simulated null data, suggesting that there is a statistically significant positive correlation between the structural and functional pairwise distances.

Nonparametric estimates of the correlation give analogous results (Spearman's $\rho$: 0.112, Kendall's $\tau$: 0.075). This is not surprising, given that the 2D scatterplot in Fig. \ref{fig:permtest} does not display an obvious nonlinear trend.

Fig. \ref{fig:why_subsampling} justifies our use of subsampling instead of bootstrapping for our confidence intervals. As we described earlier, because each bootstrap sample contains multiple copies of the same subject, the resulting structural and functional distance matrices always contain many zeros, leading to a spuriously high correlation compared to the truth. The plots show that the bootstrap distribution fails to capture the actual correlation and is significantly biased upwards, while subsampling does not have this issue because it samples from the data without replacement.

\subsection{Sparse CCA}

\begin{figure}[t]
	\centering
	\includegraphics[width=0.47\textwidth]{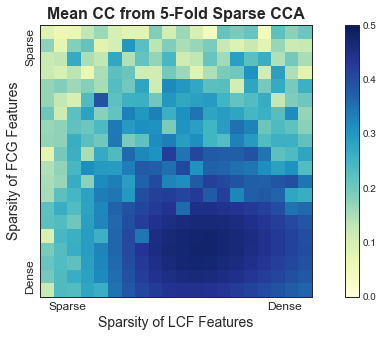}
	\includegraphics[width=0.48\textwidth]{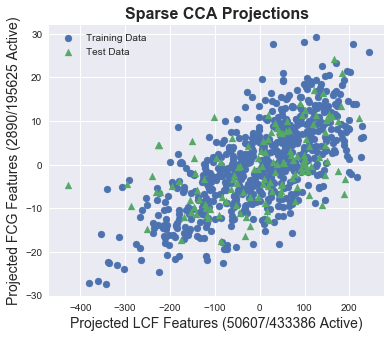}
	\caption{Cross-validated sparse CCA projections on the HCP dataset. 	
    (Left) 5-fold cross-validation plot on a validation set.
    (Right) 2D Projections of the training and test set using the best parameter found. Canonical correlations: 0.689 (train), 0.515 (test).}
    \label{fig:cv_projections_sparse_cca}
\end{figure}

For sparse CCA, we select a pair of sparsity ($\ell_1$) parameters from a 2D grid, one for LCFs and another for FCGs, that yields the maximum canonical correlation on the validation set.
Our cross-validation plot in the left panel of Fig. \ref{fig:cv_projections_sparse_cca} shows that there is a contiguous region of sparsity levels in both structural and functional features where the canonical correlation on the validation set is maximized. 
Using the optimal regularization parameters, we find that sparse CCA selects 50,607 (11.7\%) LCF features and 2,890 (1.48\%) FCG features to give a canonical correlation of 0.689 (train) and 0.515 (test).\footnote{As described in Section \ref{sec:cv}, the final training canonical correlation is computed on all 5 folds (size $\lceil{5/6}\rceil n$) of the training set. The test canonical correlation is computed on the held-out test set (size $\lfloor{1/6}\rfloor n$), which is unused during cross-validation.}

The cross-validated sparse CCA projections of the training data as well as the test data are plotted in Fig. \ref{fig:cv_projections_sparse_cca}. 
Since the objective of CCA is to maximize the correlation between these projected points, we expect to see a linearly increasing pattern in the projected space. The right panel of Fig. \ref{fig:cv_projections_sparse_cca} demonstrates this expectation: the projections of the training data and the test data exhibit similar linearly increasing patterns with a similar degree of variation. This implies that the alignment vectors we found can generalize well to unseen data in terms of correlation in the linearly projected space. 

Note that the projections still have relatively high variance across the linear trend. 
This variability is likely due to both the variance coming from the optimization problem, which is ill-conditioned and thus contains many local optima, and the variance coming from the lack of statistical consistency in the high-dimensional setting. 
Indeed, the optimal number of variables chosen by cross-validation (50,607 and 2,890) is still greater than the number of subjects (793).

\begin{figure}[t]
    \centering
    \includegraphics[width=0.48\textwidth]{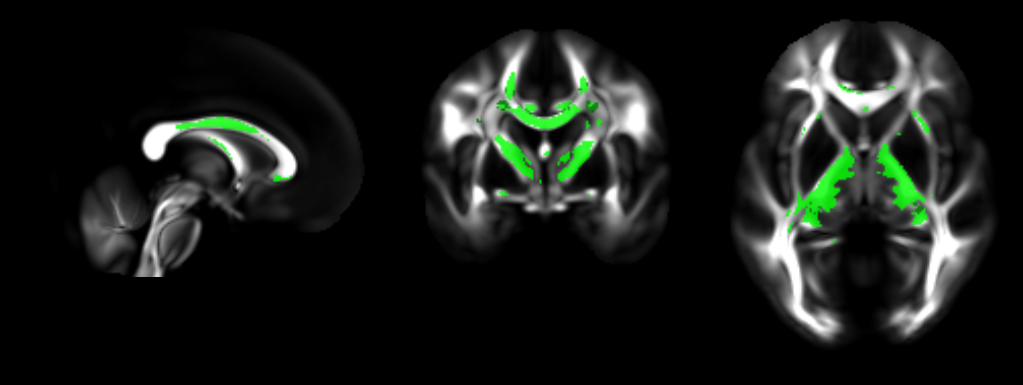}
    \includegraphics[width=0.48\textwidth]{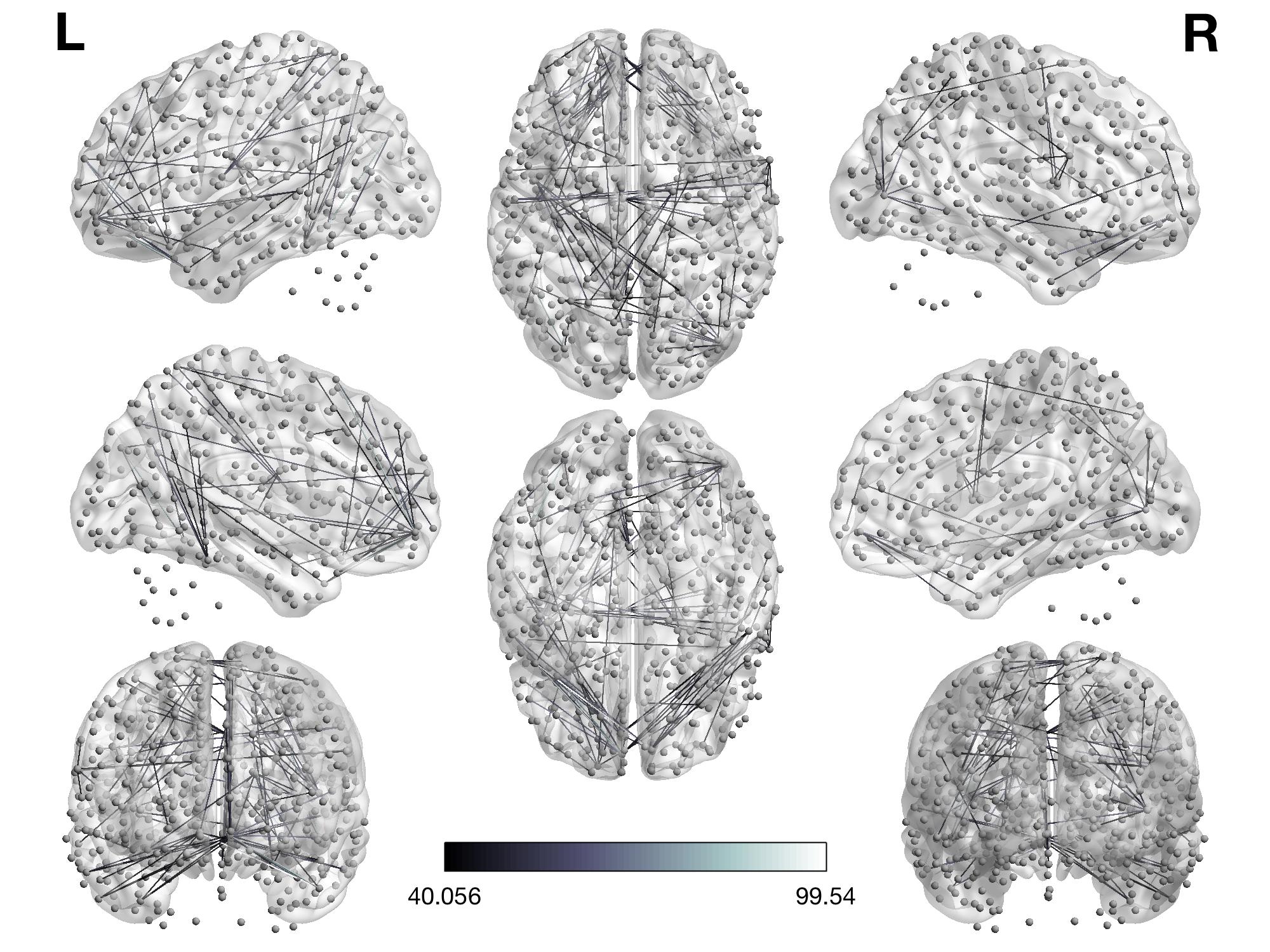}
    \caption{Visualization of cross-validated CCA projections for LCFs in MRIcron (top) and FCGs in BrainNet (bottom). We select the features that gave the best testing canonical correlation in cross-validation (best parameter, best fold).}
    \label{fig:feature_visualization}
\end{figure}

In Fig. \ref{fig:feature_visualization}, we visualize the LCF and FCG features selected by sparse CCA using the optimal sparsity parameters. 
In both modalities, sparse CCA focuses on connectivity patterns in specific regions of the brain. In particular, within the high-dimensional LCF space, the algorithm points to contiguous local pathways of the white matter structure. 
Our results show that this specific set of local white matter pathways are highly predictive of the lateral dynamics of functional connectivity between the left and right hemispheres. The structure-function association is observed between the core white matter pathways that regulate both intracortical and cortical-subcortical communication, including the corpus callosum, thalamic radiations, corticospinal, and corona radiata pathways, and the resting state functional activity in a diversity of cortical and subcortical nodes. This suggests that the structure-function relationship is strongest in the large major communication fascicles that are critical for global brain network communication.

\subsection{Canonically Correlated Subcluster Pairs}

\begin{figure*}[t]
    \centering
    \includegraphics[width=0.32\textwidth]{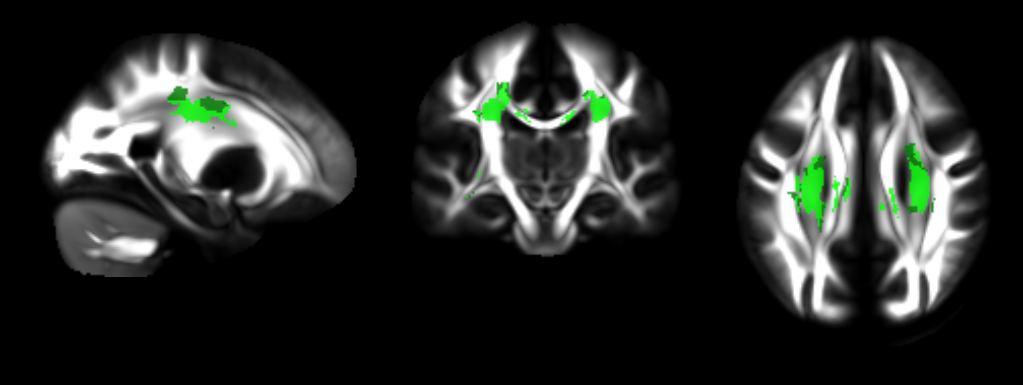}
    \includegraphics[width=0.32\textwidth]{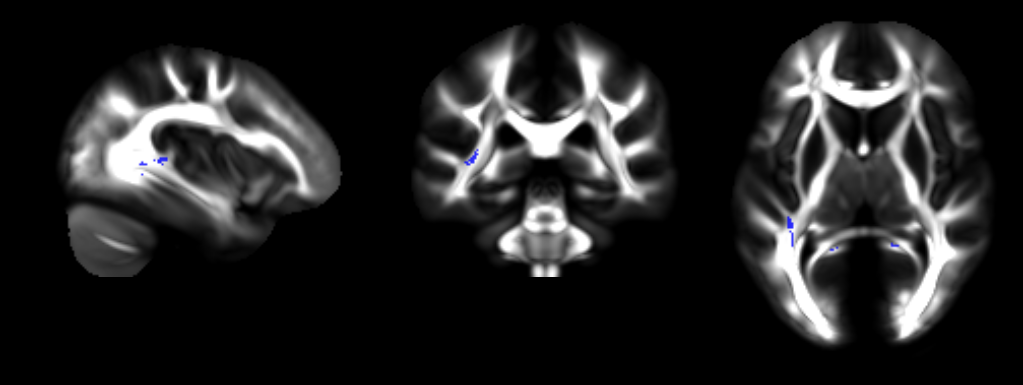}
    \includegraphics[width=0.32\textwidth]{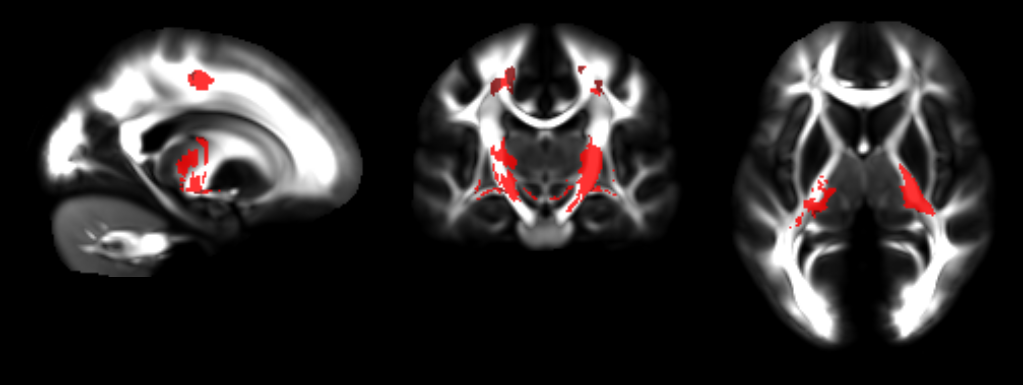}
    \includegraphics[width=0.32\textwidth]{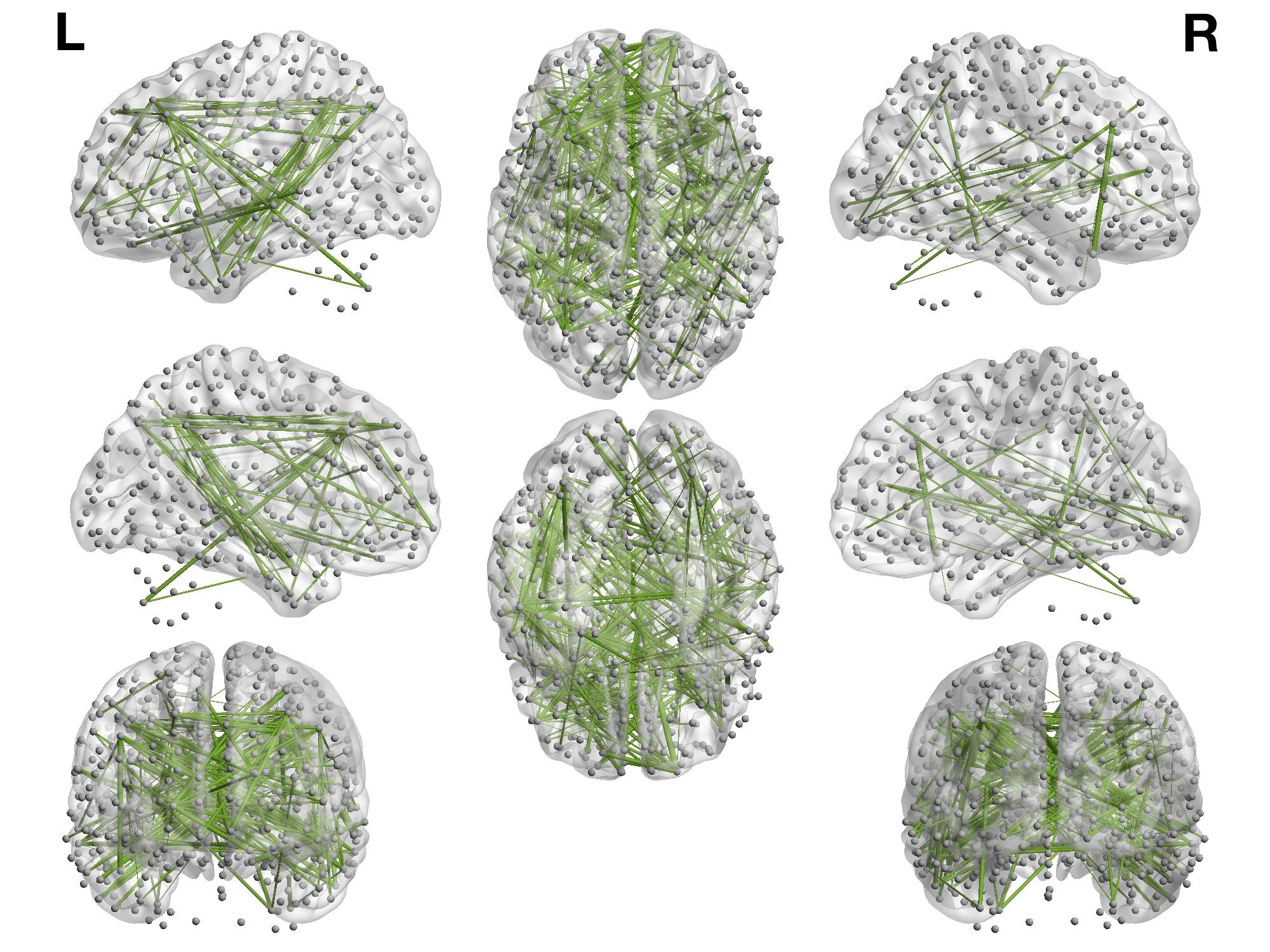}
    \includegraphics[width=0.32\textwidth]{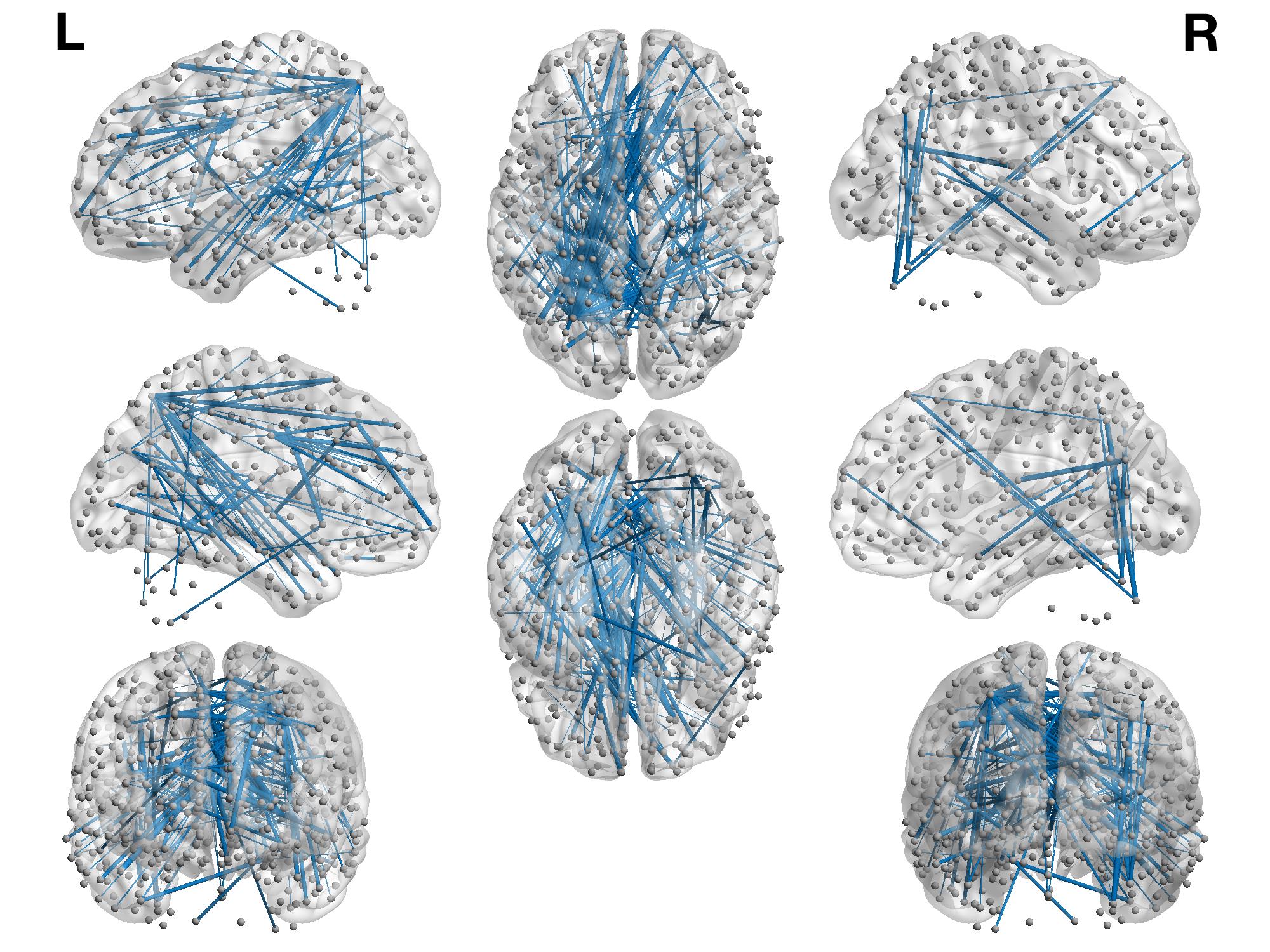}
    \includegraphics[width=0.32\textwidth]{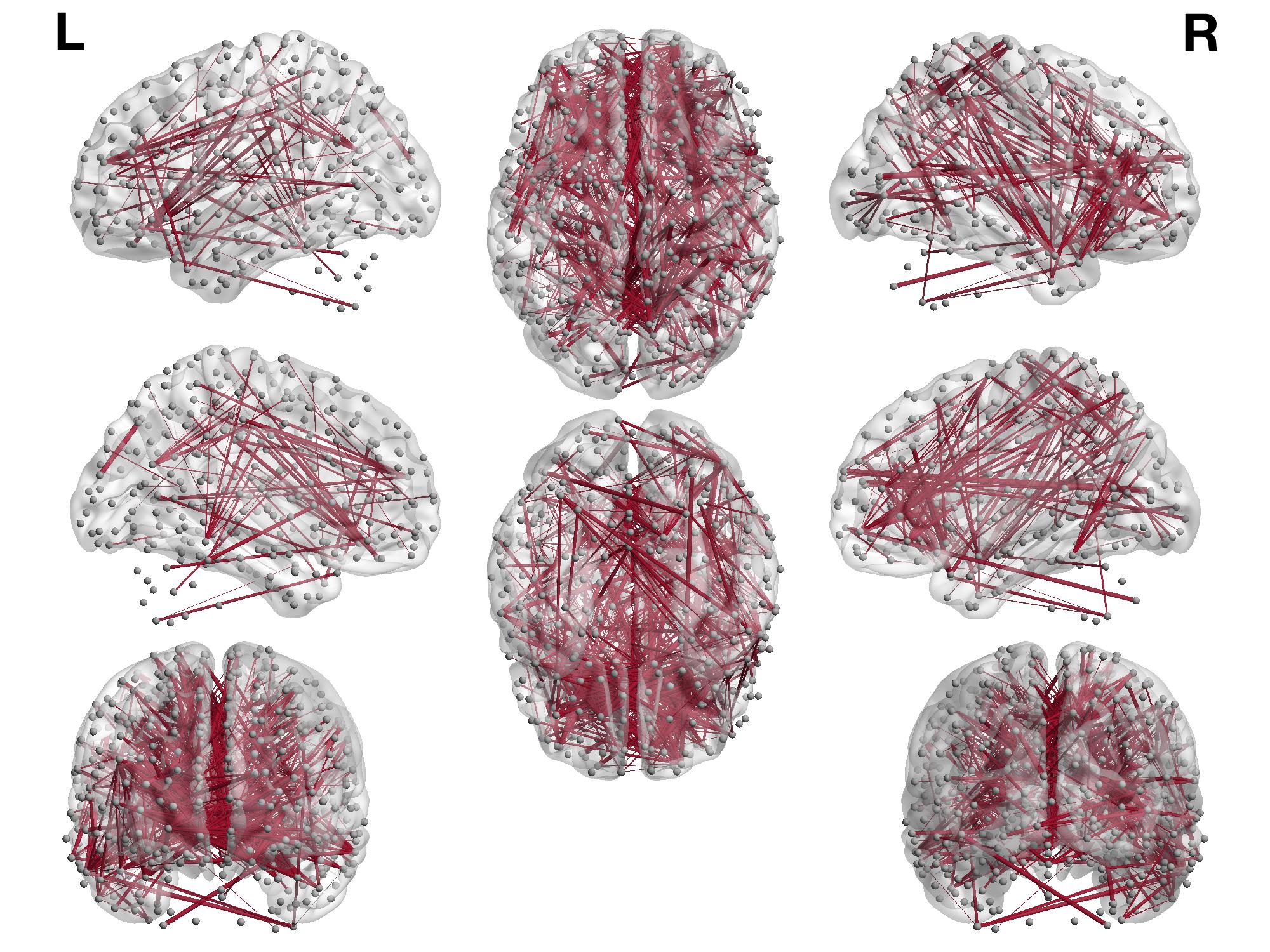}   
    \caption{Canonically correlated subcluster pairs between the selected LCF and FCG features.}
    \label{fig:subcluster_visualization}
\end{figure*}

In order to see if there is substructure in the structure-function relationships identified by sparse CCA, we decompose the canonical correlations into smaller subclusters of both the LCF and FCG entries. In Fig. \ref{fig:subcluster_visualization}, we show the three most canonically correlated pairs of LCF and FCG subclusters, which are computed by a simple agglomerative clustering (complete-linkage, same distances $\dx$ and $\dy$ respectively) of the selected LCF and FCG features into 5 subclusters each. We compute the canonical correlation between each pair of subclusters without additional regularization terms, as in \eqref{eqn:cca}.

Here we see even further specificity in the structure-function relationship. For example, variability in the centrum semiovale (Fig. \ref{fig:subcluster_visualization}, left), should predict functional dynamics of intrahemispheric and interhemispheric cortical networks. This pattern largely holds in the corresponding functional networks. In contrast, a small cluster along the inferior longitudinal fasiculus (Fig. \ref{fig:subcluster_visualization}, middle), a major means of communication along the ventral visual stream, correlates with primarily ventral visual pathway functional dynamics, as well as communication between dorsal and ventral visual streams. Finally, variability in the internal capsule (Fig. \ref{fig:subcluster_visualization}, right), a major means of communication between cortex and subcortical areas, correlates with primarily functional dynamics between cortical and subcortical nodes. Thus, the specificity of the structure-function relationships identified in this subclustering analysis is consistent with \textit{a priori} predictions derived from the neuroanatomical literature.

\section{Discussion}

In this paper, we show how variability in local white matter architecture is associated with global patterns of functional brain dynamics. Using distance-based correlations, we found a small, but significant effect whereby individuals with more similar local white matter architecture tended to also be more similar in their functional connectome. Using sparse CCA approaches, we were able to show that individual variability in white matter architecture along major brain fascicles correlated with individual differences in functional dynamics within the specific class of brain networks that would be predicted by existing neuroanatomical knowledge. Thus, in conjunction with the constraints of global end-to-end structural connectivity \cite{hermundstad2013structural}, our results highlight how variability in the local white matter systems also impacts global brain communication.

\section*{Acknowledgments}
The research was sponsored by the U.S. Army Research Laboratory, including work under Cooperative Agreement Number W911NF-10-2-0022, and the views espoused are not official policies of the U.S. Government.
S.B was partially supported by the NSF grant DMS-1713003. 

\clearpage

\bibliographystyle{IEEEtran}
\bibliography{references}

\end{document}